\documentclass[aps,prl,reprint,superscriptaddress,floatfix,nofootinbib]{revtex4-2}
\usepackage[usenames, dvipsnames]{color}
\usepackage{amsmath}
\usepackage{amsfonts}
\usepackage{comment}
\usepackage{hyperref}
\usepackage{ragged2e}
\usepackage[caption=false]{subfig}
\hypersetup{colorlinks=true, linkcolor=Blue, citecolor=Blue, filecolor=Blue, urlcolor=Blue}
\usepackage{graphicx}
\usepackage{physics}
\usepackage[normalem]{ulem}

\newcommand{\rmv}[1]{{\color{red}\sout{#1}}}

\begin{document}
\title{Topological chirality of dissipative limit cycles in an open Dicke model}
\author{Nikolay Yegovtsev}
\email{nyegovts@purdue.edu}
\affiliation{Department of Physics and Astronomy and IQ Initiative, University of Pittsburgh, Pittsburgh, Pennsylvania 15260, USA}
\affiliation{Department of Physics and Astronomy, Purdue University, West Lafayette, Indiana 47907, USA}
\author{Sayan Choudhury}
\email{sayanchoudhury@hri.res.in}
\affiliation{Harish-Chandra Research Institute, Chhatnag Road, Jhunsi,
Prayagraj 211 019, India}
\affiliation{Homi Bhabha National Institute, Training School Complex,
Anushakti Nagar, Mumbai 400 094, India}
\author{W. Vincent Liu}
\email{wvliu@pitt.edu}
\affiliation{Department of Physics and Astronomy and IQ Initiative, University of Pittsburgh, Pittsburgh, Pennsylvania 15260, USA}

\date{\today}
	
\begin{abstract}

In an open, $U(1)$-symmetric Dicke model with chiral atom-cavity couplings, we show that dissipation drives two limit-cycle phases of opposite chirality in the thermodynamic limit, obtaining exact analytical solutions. These phases are separated by a $U(1)$-broken superradiant state, lending the phase diagram a topological character, and persist under $U(1)$-preserving perturbations, making them candidates for chiral continuous time crystals. In addition to stable normal and inverted steady states, the model also exhibits multistability, where the long-time dynamics is set by the initial state. Our results establish dissipation as a resource for inducing chiral dynamical order in light-matter coupled systems.
\end{abstract}

\maketitle

Characterizing equilibrium phases of matter and the transitions between them, both thermal and quantum, lie at the heart of condensed matter physics~\cite{Sachdev_2011,Nishimori_2010,carollo2020geometry}. With the advent of controllable quantum systems, these investigations have now been extended to non-equilibrium settings~\cite{moessner2017equilibration,abanin2019colloquium,zaletel2023colloquium}. In this context, dissipation has emerged as a powerful mechanism for realizing and stabilizing novel non-equilibrium phases of matter~\cite{harrington2022engineered,mi2024stable}, ranging from time crystals~\cite{kessler2021observation,kongkhambut2022observation,wu2024dissipative,liu2025bifurcation} to mixed-state topological order~\cite{fan2024diagnostics,wang2025intrinsic,ellison2025toward,sohal2025noisy}, thereby playing a crucial role in the development of quantum technologies~\cite{montenegro2023quantum,sannia2024dissipation}.

A paradigmatic model, where the effect of dissipation has been extensively investigated, is the open Dicke model (DM), which describes the dynamics of $N$ two-level atoms interacting with a quantized cavity mode~\cite{DickeOG,garraway2011dicke,kirton2019introduction,frisk2019ultrastrong}. This system exhibits a dissipative $\mathbb{Z}_2$ symmetry-breaking transition from the normal to the superradiant phase when the strength of the atom-photon coupling $\lambda$ exceeds a critical value, $\lambda_c$~\cite{HeppLieb, WangHioe,dimer2007proposed}. This phase transition is manifested as a dynamical phase transition, a pitchfork bifurcation, from one attractor to two attractors~\cite{klinder2015dynamical,roses2020dicke}. Intriguingly, in generalized DMs, even an infinitesimal amount of dissipation can lead to a significantly richer phase diagram, compared to their closed counterparts~\cite{baksic2014controlling,soriente2018dissipation,keeling2010collective, bhaseen2012, soriente2018dissipation,unbalanceddicke2020}. Crucially, however, the superradiant phases in these generalized open DMs are characterized by the spontaneous breaking of a discrete $\mathbb{Z}_2$ symmetry, while dissipation obstructs the realization of a continuous $U(1)$ symmetry-broken superradiant phase~\cite{soriente2018dissipation}. This raises two important questions: (a) Is it possible to realize $U(1)$-broken phases of matter in the presence of dissipation, and (b) can such phases exhibit chirality?

In this Letter, we answer these questions affirmatively by providing a route to realize chiral $U(1)$-broken limit-cycle (LC) phases and superradiant steady states in a dissipative two-mode $U(1)$-symmetric DM dubbed the open chiral DM. This model describes the dynamics of $N$ two-level systems (TLS) interacting with two degenerate cavity modes~\cite{yegovtsev2026robustcontinuoussymmetrybreaking}, and it hosts a robust $U(1)$-broken superradiant phase in the absence of dissipation. The addition of dissipation endows this system with a much richer phase diagram. Notably, we analytically demonstrate that the system hosts two distinct LC phases, characterized by opposite chiralities, separated by the $U(1)$-broken superradiant phase. We further establish that the chiral LC phases persist in the presence of $U(1)$ symmetry-preserving perturbations, a robustness that we argue makes them appealing candidates for chiral time crystals. Finally, we show that this system can exhibit multistability, where the long-time dynamics is determined by the initial state. The analytic tractability of these results stems from a structural feature of the model: although the chiral symmetry is only weakly respected by the dissipation, every dynamical phase balances the photon
fluxes in the two cavity modes, which in turn permits closed-form solutions. Our results establish a path for realizing novel chiral dynamical phases of matter in atom-photon coupled systems.

{\emph {Model:}}  The chiral Dicke model is the many-atom generalization of the chiral Rabi model~\cite{chiral2019} and describes the dynamics of $N$ TLS coupled to the two degenerate cavity modes through co-rotating and counter-rotating terms with strengths $g_1$ and $g_2$~\cite{yegovtsev2026robustcontinuoussymmetrybreaking}:
\begin{equation}
\label{eq:Hcdm}
\begin{split}
H_{\rm CDM} = &\, \omega_c(a_1^\dagger a_1+a_2^\dagger a_2)+\omega_zS^{z}\\
& +\frac{g_1}{\sqrt{N}}(a_1S^{+}+a_1^\dagger S^{-}) +\frac{g_2}{\sqrt{N}}(a_2S^{-}+a_2^\dagger S^{+}),   
\end{split}
\end{equation}
where the collective spin operator, $S^{\pm} = \frac{1}{2} \sum_{j=1}^N (\sigma_j^{\rm x} \pm i \sigma_j^{\rm y})$ and $\{\omega_c,\omega_z,g_1,g_2\} >0$. 
\begin{figure*}[t]
        \includegraphics[width = \textwidth]{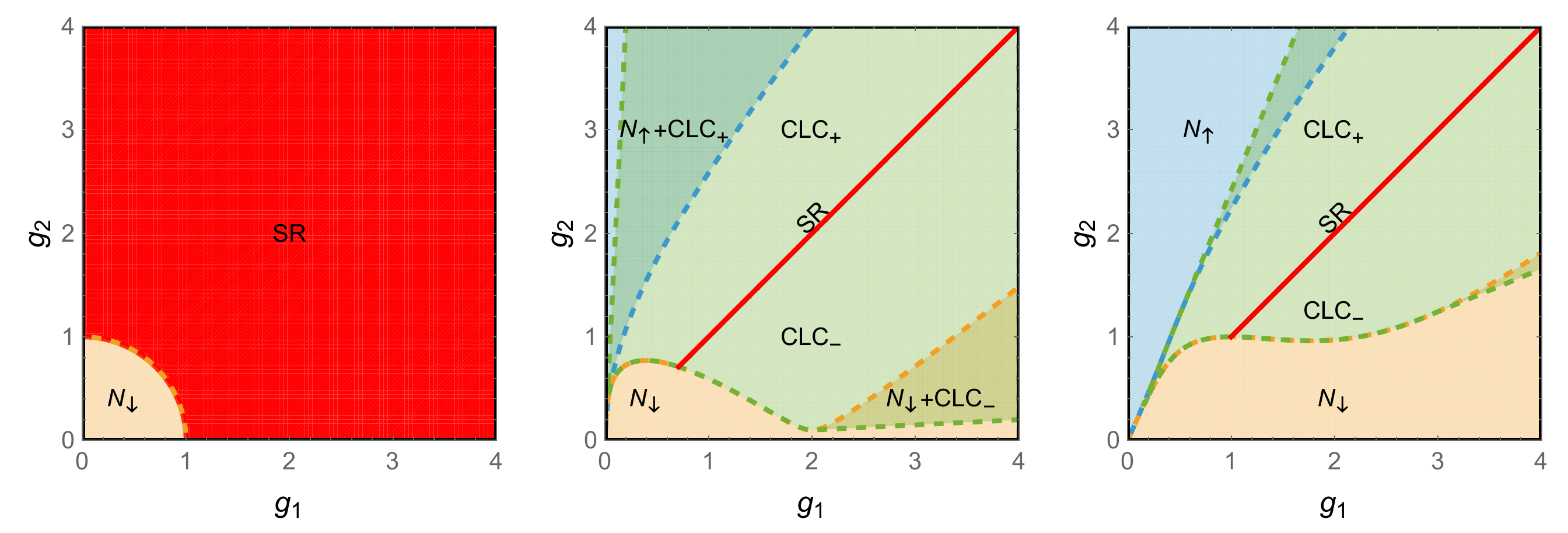}
    \caption{Phases of the chiral Dicke model for the choice of parameters $\omega_c=\omega_z=1$. The leftmost figure is the ground state phase diagram of the closed system $\kappa=0$. The middle and the rightmost figures are the dynamical phase diagram in the presence of small $\kappa=0.1$ and moderate $\kappa=1$ dissipation respectively. The orange and red regions on the ground state phase diagram correspond to the $U(1)$-symmetric normal $N_{\downarrow}$ and $U(1)$-broken superradiant phase $SR$. On the dynamical phase diagram, the same colors correspond to the stable normal and the superradiant steady states respectively. There are additional regions on the dynamical phase diagram corresponding to the stable inverted state $N_{\uparrow}$ (blue) and chiral limit cycle phases ${\rm CLC}_{\pm}$ (green). The subscript of chiral phases denotes the sign of the frequency, $\omega$ (Eq.~\eqref{eq:omega}) and thereby characterizes the chirality. The intersections of the green region with the orange and blue ones correspond to the regions of multistability. The phase boundaries are represented by the dashed lines, and obtained using linear stability analysis~\cite{SuppMat} and Eqs.~\eqref{eq:omega} and \eqref{eq:lcz}.}
    \label{fig:1}
\end{figure*}
The defining feature of this model is the chiral $U(1)$ symmetry: $a_1\to e^{i\phi}a_1$, $a_2\to e^{-i\phi}a_2$, $S^{\pm} \to e^{\mp i\phi}S^{\pm}$, which in turn leads to the conservation of the total angular momentum:
\begin{equation}
L^z = a_1^\dagger a_1-a_2^\dagger a_2+S^z.    
\end{equation}
We introduce the effect of dissipation by employing the Lindblad equation:
\begin{equation}
    \dot{\rho} = -i[H,\rho] + \sum_{i}\kappa_i\left(2L_i \rho L_i^\dagger - \{L_i^\dagger L_i, \rho\}\right)
\end{equation}
with the jump operators $L_1 = a_1$ and $L_2 = a_2$ and $\kappa_1=\kappa_2=\kappa$. We investigate the dynamics of this system by solving the equations of motion $\dot{\hat{O}} =i[H,\hat{O}]+\sum_i\kappa_i\left(2L_i^{\dagger} \hat{O}L_i-\{L_i^\dagger L_i,\hat{O}\}\right)$, where $\hat{O}$ are the operators of interest. We observe that while the Hamiltonian and Liouvillian dynamics respect the $U(1)$ symmetry, the jump operators themselves carry a nonzero 
$U(1)$ charge and therefore do not commute with $L^z$. Consequently, this chiral $U(1)$ symmetry constitutes a weak symmetry~\cite{lessa2025strong,gu2025spontaneous} and $\dot{L}^z = -2\kappa(a_1^\dagger a_1-a_2^\dagger a_2)$ is no longer identically zero. Crucially, however, we demonstrate that all the dynamical phases of this model satisfy $\dot{L}^z=0$, despite the non-conservation of $L^z$. We now obtain the phase diagram of this model in the thermodynamic limit.



\emph{Mean-field phase diagram:}  Due to the collective nature of the chiral DM, a mean-field treatment of this system becomes exact in the thermodynamic limit. To this end, we obtain the equations of motion for $\langle a_1\rangle = \alpha_1$, $\langle a_2 \rangle=\alpha_2$, $\langle S^-\rangle = s$, $\langle S^z\rangle = z$ employing standard mean-field decoupling:

\begin{equation}
\label{eq:eom_mf}
\begin{split}
&\dot{\alpha}_1 = -(\kappa+i\omega_c)\alpha_1- \frac{ig_1}{\sqrt{N}}s,\\
&\dot{\alpha}_2 = -(\kappa+i\omega_c)\alpha_2- \frac{ig_2}{\sqrt{N}}s^*,\\
&\dot{s} = -i\omega_zs +\frac{2i}{\sqrt{N}}(g_1\alpha_1+g_2\alpha_2^*)z,\\
&\dot{z} = \frac{ig_1}{\sqrt{N}}\left(-\alpha_1s^*+\alpha_1^* s\right)+\frac{ig_2}{\sqrt{N}}\left(\alpha_2s-\alpha_2^* s^*\right).
\end{split}    
\end{equation}

We identify the steady states of this system by setting the left-hand side of Eq.~\eqref{eq:eom_mf} to zero. There are two trivial steady states - $\alpha_{1ss}=\alpha_{2ss}=s_{ss}=0$ and $z_{ss} = \mp N/2$ - that correspond to the normal and inverted states. We identify the superradiant steady state by obtaining $\alpha_1$ and $\alpha_2$ from the first two equations and substituting them into the last equation:
\begin{equation}
\label{eq:fluxbalance}
\dot{z} = \frac{2\kappa\,|s|^2\left(g_2^2-g_1^2\right)}{N(\kappa^2+\omega_c^2)}
        = 2\kappa\left(|\alpha_2|^2-|\alpha_1|^2\right).
\end{equation}
Consequently, the superradiant steady state exists with $s\ne 0$ only when the photon fluxes are balanced ($|\alpha_1| = |\alpha_2|$), leading to the condition $g_1=g_2=g$, when $\kappa \ne 0$. At $\kappa = 0$, $\dot{z}$ is identically zero, and no such constraint exists. Furthermore, employing the equation for $\dot{s}$, we obtain:
\begin{equation}
\label{eq:zss}
z_{ss}^{\rm SR} = -\frac{\omega_zN(\kappa^2+\omega_c^2)}{4\omega_c g^2} = -\frac{N}{2}\frac{g_c^2}{g^2},   
\end{equation}
where $z_{ss}^{\rm SR} \geq -\frac{N}{2}$ and $g_c = \sqrt{\omega_z(\kappa^2+\omega_c^2)/(2\omega_c)}$, which implies that the superradiant state is possible only when $g \geq g_c$. The expressions for the remaining quantities read:
\begin{equation}
\label{eq:sr}
s^{\rm ss}  = \mathcal{A} e^{i\phi},\,\, \alpha_{1}^{\rm ss} =  \mathcal{B} \exp[i\phi],\,\,
\alpha_{2}^{\rm ss} = \mathcal{B} \exp[-i\phi],    
\end{equation}
where 
\begin{equation}
    \mathcal{A}=\frac{N}{2}\sqrt{1-(g_c/g)^4}, \, \mathcal{B} = -\frac{ig\sqrt{N}}{2(\kappa+i\omega_c)}\sqrt{1-(g_c/g)^4},\nonumber
\end{equation} 
and we have used the conservation of the total spin $z^2+|s|^2=N^2/4$. The value of $\phi$ is arbitrary, signifying the spontaneous breaking of the chiral $U(1)$ symmetry. We establish the phase boundaries by performing a linear stability analysis for the aforementioned steady states in the regime of weak ($\kappa=0.1$) and moderate ($\kappa=1$) dissipation~\cite{SuppMat}; we set $\omega_c=\omega_z=1$ for these computations. 

The resulting phase diagram (Fig.~\ref{fig:1}) exhibits several interesting features. We begin by noting that dissipation modifies the phase diagram drastically from its closed counterpart~\cite{yegovtsev2026robustcontinuoussymmetrybreaking}. While in the latter, the normal-to-superradiant phase transition occurs along the circle parametrized by $\sqrt{g_1^2+g_2^2} = g_c^{(0)} = \sqrt{\omega_z\omega_c}$, an arbitrarily weak dissipation collapses this entire critical circle onto the single line $g_1=g_2$. Furthermore, the introduction of dissipation primarily destroys the $U(1)$-broken phase in the Tavis-Cummings ($g_1\gg g_2$) and the anti-Tavis-Cummings ($g_1\ll g_2$) regimes, in agreement with the results reported in Ref.~\cite{soriente2018dissipation}. Finally, while the $U(1)$-broken superradiant steady state exists along one line, $U(1)$-breaking occurs over a much larger parameter regime, where chiral limit cycles are formed. The chiral DM also exhibits multistability, where different initial states flow to different attractors, akin to other generalized DMs~\cite{keeling2010collective, bhaseen2012, unbalanceddicke2020}. We analyze these dynamical features next.

\begin{figure}[t]
        \includegraphics[width = 0.4\textwidth]{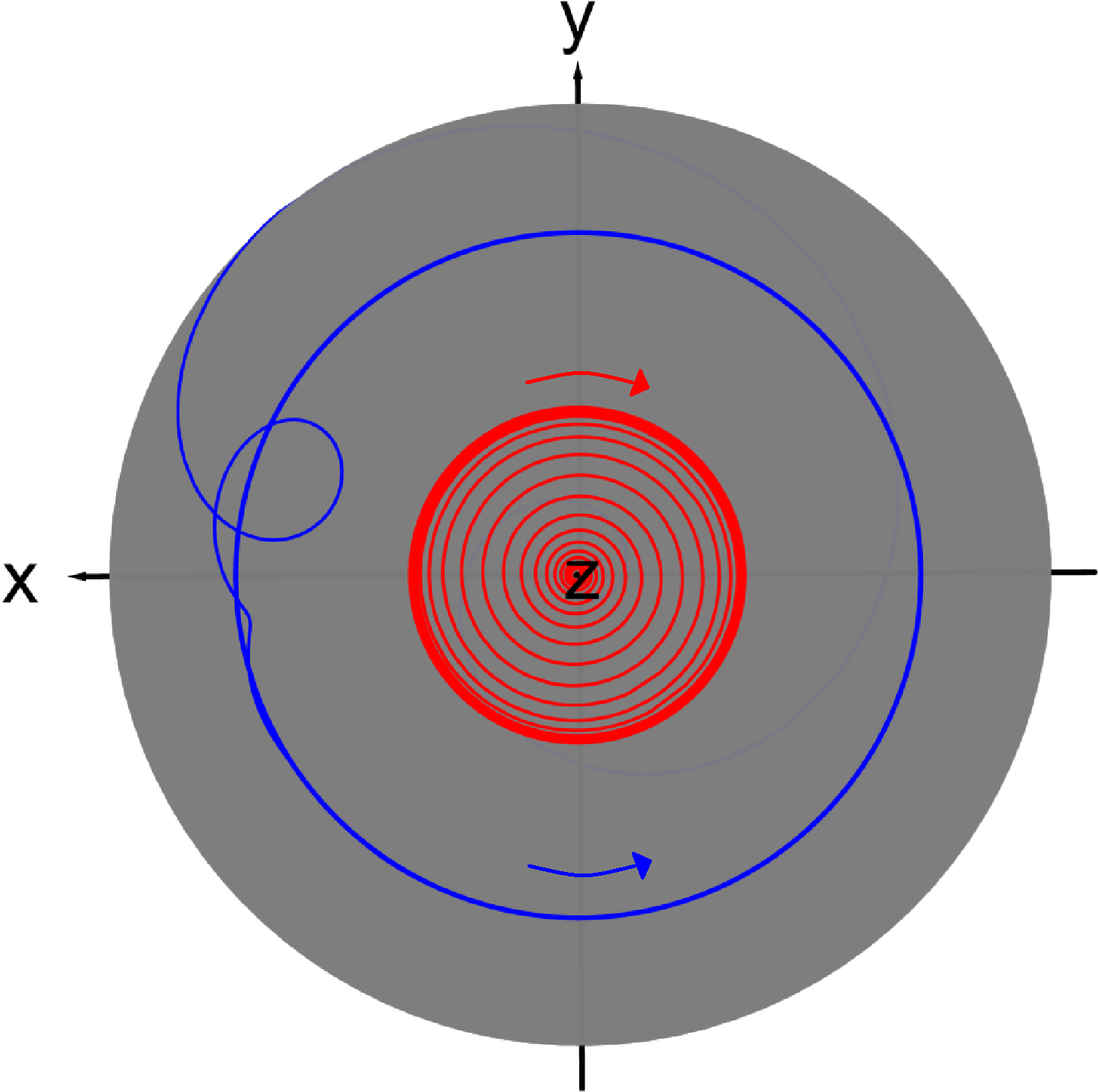}
    \caption{Limit cycles in the chiral Dicke model for the choice of parameters $\omega_c=\omega_z=\kappa=1$ as seen from the south pole of the collective Bloch sphere. The limit cycle in red corresponds to the choice $g_1=1.5$, $g_2=1$ and the limit cycle in blue to $g_1=1$, $g_2=1.15$. The initial state is chosen in the vicinity of the south pole (red) or the north pole (blue) of the collective Bloch sphere. The chirality of the limit cycles is different, as indicated by arrows of the corresponding color, in agreement with the results of Eq.~\eqref{eq:lcansatz} and Eq.~\eqref{eq:omega}.}
    \label{fig:2}
\end{figure}

\emph{Limit cycle phase:} Before proceeding to discuss the $U(1)$-broken limit cycle (LC) phases that emerge in the chiral DM, we note that $\mathbb{Z}_2$-broken LCs have been numerically observed both in the unbalanced DM~\cite{unbalanceddicke2020} and its $U(1)$-extension~\cite{KeelingU1}. An analytical characterization of these $\mathbb{Z}_2$-broken LCs has proven to be difficult, since the motion of the spin on the collective Bloch sphere occurs primarily on some general plane intersecting the sphere. Intriguingly, for the present case of the chiral DM, the LCs correspond to the spontaneous breaking of the chiral $U(1)$ symmetry, leading to oscillations of $s = x-iy$ only. Thus, the LCs occur parallel to the $xy$ plane at a fixed value of $z$ on the collective Bloch sphere, as shown in Fig.~\ref{fig:2}, thereby enabling us to obtain a simple analytic solution for the LCs. 

Motivated by the numerical analysis, we seek solutions of Eq.~\eqref{eq:eom_mf} in the form:
\begin{equation}
\label{eq:lcansatz}
z = \frac{N}{2}\cos(\theta), \hspace{5mm} s = \frac{N}{2}\sin(\theta)e^{-i\omega t+i\phi},   
\end{equation}
where $\theta$ is the standard angle that parametrizes the $z$-axis in the spherical coordinates, and $\phi$ is an arbitrary phase associated with the $U(1)$-breaking. Substituting this into the equations for $\dot{\alpha}_1$ and $\dot{\alpha}_2$, and neglecting transient terms, we obtain:
\begin{equation}
\label{eq:lca1a2}
\begin{split}
&\alpha_1(t) = \frac{\sqrt{N}g_1\sin(\theta)}{2}\frac{e^{-i\omega t+i\phi}}{(\omega -\omega_c)+i\kappa},\\
&\alpha_2(t)= -\frac{\sqrt{N}g_2\sin(\theta)}{2}\frac{e^{i\omega t-i\phi}}{(\omega +\omega_c)-i\kappa}.    
\end{split}    
\end{equation}
Substituting those expressions into the equation for $\dot{z}$ yields:
\begin{equation}
\label{eq:zlcycle}
\frac{g_1^2}{(\omega-\omega_c)^2+\kappa^2}-\frac{g_2^2}{(\omega +\omega_c)^2+\kappa^2} =  0.  \end{equation}
Interestingly, this is the generalization of the photon flux balance condition in Eq.~\eqref{eq:fluxbalance} to finite $\omega$, leading to $\dot{z}=0$. Furthermore, the time-independence of $n_1= \vert \alpha_1 \vert^2$ and $n_2 = \vert \alpha_2 \vert^2$ leads to the effective conservation of $L^z$ at long times.

The flux balance condition yields the LC oscillation frequency:
\begin{equation}
\label{eq:omega}
\omega =\frac{-g_{+} (\omega_c-\sqrt{\tilde{\omega}^2})}{g_{-}},  
\end{equation}
where $\tilde{\omega}^2 = \omega_c^2-(\kappa^2+\omega_c^2)(g_{-}/g_{+})^2,\, g_{\pm} = g_1^2\pm g_2^2$, and the sign in front of the square root guarantees that $\omega=0$ when $g_1=g_2$, yielding the superradiant steady state; this result does not depend on $\omega_z$. The two LC phases are distinguished by the directionality of the collective spin rotation, $\mathcal{C} = {\rm sign}[\omega]$, and we label them ${\rm CLC}_{+}$ and ${\rm CLC}_{-}$ for $\mathcal{C}=1$ and $-1$ respectively. Since the numerator of Eq.~\eqref{eq:omega} is always negative and real whenever $\tilde{\omega}^2 \ge 0$, it follows that $\mathcal{C} = -{\rm sign}[g_{-}]$. Consequently, the ${\rm CLC}_{+}$ and ${\rm CLC}_{-}$ phases exist when $g_1<g_2$ and $g_1>g_2$ respectively. The chirality of the LC can also be probed experimentally from Eq.~\eqref{eq:lca1a2}, where the two photon modes oscillate at opposite frequencies, so that $\mathcal{C}$\rmv{,}{} can be extracted from a frequency-resolved heterodyne measurement of the cavity output~\cite{Ferri2021,Finger2024}.

Notably, this feature of two chiral limit cycles persists beyond the symmetric regime. It survives in a more general chiral DM with different values of $\omega_1, \omega_2$ and $\kappa_1, \kappa_2$, and it survives under $U(1)$-preserving perturbations of the form $U_1(a_1^\dagger a_1+a_2^\dagger a_2)/N$ and $U_2(a_1^\dagger a_1-a_2^\dagger a_2)/N$~\cite{SuppMat}. In these cases, the superradiant state is displaced to a different line in the $(g_1,g_2)$ plane. The resulting phase diagram is topological in character: the chirality $\mathcal{C}$ cannot change sign without passing through the $\mathcal{C}=0$ superradiant state, since $\omega$ varies continuously with $g_1,g_2$ and can therefore only cross zero rather than jump between $\pm1$. The superradiant state thus plays the role of a defect separating two regions of opposite charge, retaining a nonzero order parameter $|s|$. Together with the fact that these LCs sustain persistent oscillations in the thermodynamic limit, this robustness makes them promising candidates for chiral time crystals. However, in order to establish the time-crystalline nature, it is crucial to characterize the dependence of the oscillation lifetime on the number of atoms~\cite{iemini2018boundary,wang2025boundary}; this analysis is beyond the scope of this work.

\begin{figure}[t]
        \includegraphics[width = 0.4\textwidth]{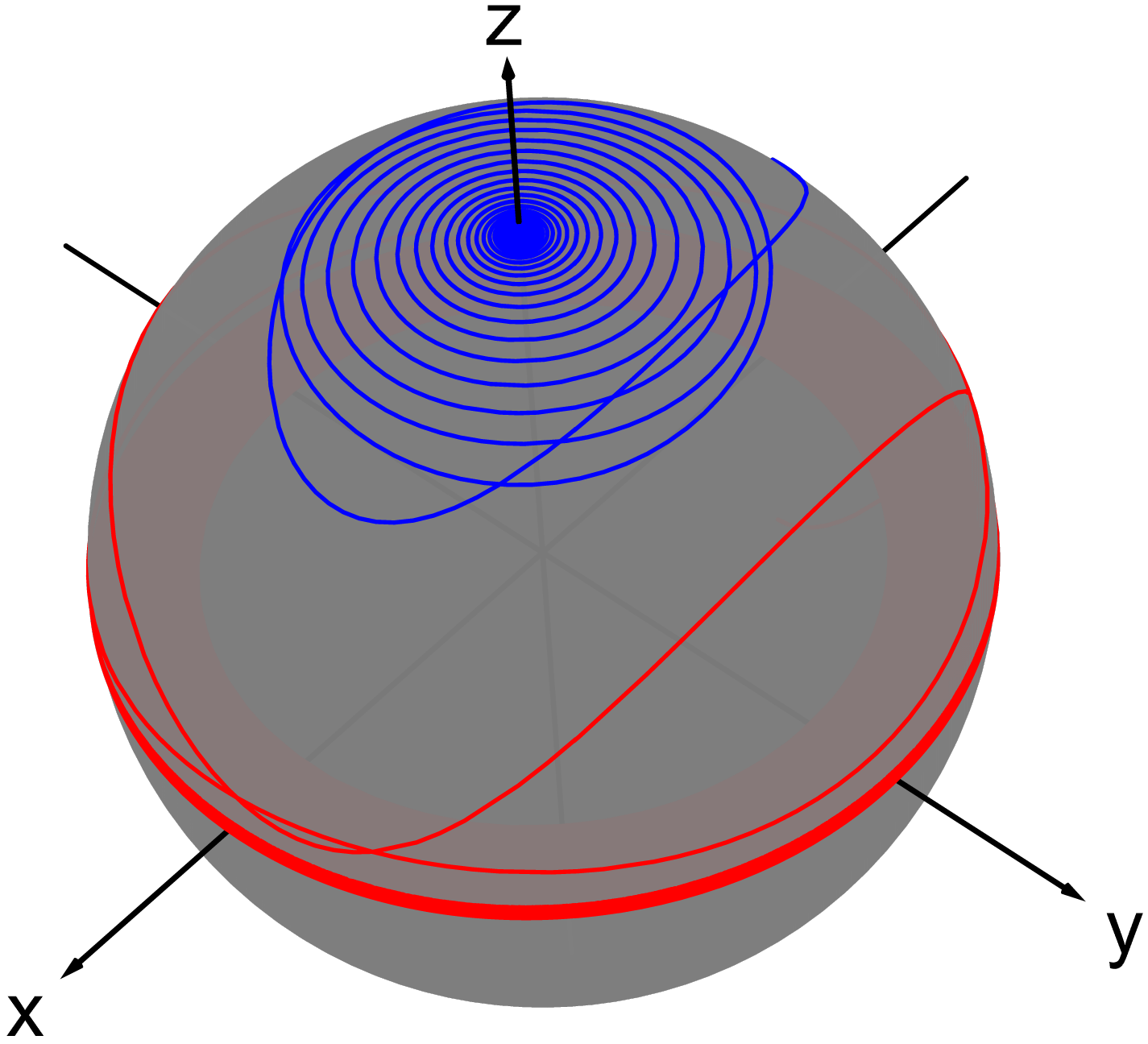}
    \caption{Multistability for  $\omega_c=\omega_z=\kappa=1$, $g_1=1.25$ and $g_2=2.8$. The initial state prepared in the southern hemisphere flows to the limit cycle phase ${\rm CLC}_{+}$, while the state prepared in the northern hemisphere flows to the stable inverted state $N_{\uparrow}$.}
    \label{fig:3}
\end{figure}

We complete the characterization of the LC phases in the thermodynamic limit by computing the value of $z$ corresponding to the LC, $z^{\ast}$ by substituting the results obtained above into the equation for $\dot{s}$:
\begin{equation}
\label{eq:lcz}
z^{\ast} =  \frac{N(\omega-\omega_z)}{2\omega_c}\left[\frac{g_1^2}{(\omega-\omega_c)^2+\kappa^2}+\frac{g_2^2}{(\omega+\omega_c)^2+\kappa^2}\right]^{-1},
\end{equation}
such that $\theta$ in Eq.~\eqref{eq:lcansatz} takes the value $\cos^{-1}(2z^{\ast}/N)$. As $z^{\ast} \rightarrow z_{ss}^{\rm SR}$ when $\omega \rightarrow 0$, thereby demonstrating that the superradiant state serves as the zero-frequency limit of the LC family. We conclude that the LC phases exist in the parameter regime characterized by $\tilde{\omega}^2\ge 0 $ and $-N/2<z<N/2$. The maximum LC frequency that can be attained is obtained by setting $\tilde{\omega} = 0$: $|\omega_{\rm max}| = |g_{+}\omega_c/g_{-}| = \sqrt{\kappa^2+\omega_c^2}$, a bound set by the cavity scales alone. For $\tilde{\omega}^2<0$, the frequency acquires an imaginary part, so sustained LC oscillations are no longer possible. In Fig.~\ref{fig:1} the LC regions also intersect the regions of stability of the normal or inverted state, leading to multistable dynamical regimes, where different initial states flow to different attractors (Fig. \ref{fig:3}). Finally, while Fig.~\ref{fig:2} and  Fig.~\ref{fig:3} correspond to specific parameter choices, we have verified that trajectories initialized at generic points within the LC regions in Fig.~\ref{fig:1} flow onto the predicted LCs. We can gain further insight into the nature of these LCs through an atom-only framework, which we discuss next.

\emph{Atom-only theory:} In order to gain further insight into the atomic dynamics, we construct an atom-only theory in the regime of large dissipation and cavity frequency, $\omega_z\ll \kappa,\omega_c$, by eliminating the cavity modes. One caveat is that the standard adiabatic elimination procedure results in conservative spin dynamics rather than dissipative, thereby failing to capture the long-time physics~\cite{atomonlyz2}. This issue has been addressed in Refs.~\cite{atomonlyz2, atomonlyJager} for the standard DM using advanced techniques. However, the correct behavior can be readily obtained from the semiclassical equations by performing adiabatic elimination that includes the next-to-leading order corrections to the cavity field~\cite{SuppMat}. We generalize this analysis and derive an effective $U(1)$ atom-only theory that qualitatively captures the nature of the atomic dynamics of the chiral DM~\cite{SuppMat}. This procedure provides a complementary route to the fourth-order Keldysh-Redfield approach to obtain $U(1)$ atom-only theories~\cite{atomonlyu1}. The collective spin dynamics is governed by:
\begin{equation}
\label{eq:spindyn}
\dot{\mathbf{S}} = \{\mathbf{S},H\} - \Gamma_1\mathbf{S}\times(\mathbf{S}\times \hat{z}),    
\end{equation}
where the first term corresponds to the Poisson bracket, and the second to the dissipative terms. The exact expressions for the Hamiltonian $H$ and $\Gamma_1$ are presented in the Supplemental Material~\cite{SuppMat}. Unlike the standard DM, which maps to the corresponding Lipkin-Meshkov-Glick Hamiltonian in the atom-only theory, here we must also include the $(S^z)^3$ term. 

This atom-only description provides an alternative lens for viewing the origins of the LCs, since the two terms in Eq.~\eqref{eq:spindyn} steer the collective spin in orthogonal directions on the Bloch sphere. On the one hand, $H$ is a function of $S^z$ and $\mathbf{S}^2$ alone, and generates an azimuthal precession of the collective spin at fixed $z$. On the other hand, the dissipative term is polar in nature, and drives the system along lines of constant azimuth towards one of the poles at a rate:
\begin{equation}
\label{eq:zdot_atomonly}
\dot{z} = \Gamma_1\left(\mathbf{S}^2-z^2\right) = \Gamma_1|s|^2,
\end{equation}
which has the same form as Eq.~\eqref{eq:fluxbalance}. The spin is thus driven towards the normal (inverted) state at the south (north) pole when $\Gamma_1<0$ ($\Gamma_1>0$). Crucially, $\Gamma_1$ depends linearly on $S^z$~\cite{SuppMat}, and the polar motion ceases at $z=z^{\ast}$ at which $\Gamma_1$ vanishes. The dissipative rate $\Gamma_1$ is expressed as $A+BS^z$~\cite{SuppMat}, so that $\Gamma_1 = B(z-z^{\ast})$ and
\begin{equation}
\frac{d}{dt}\left(z-z^{\ast}\right)^2 = 2B\left(z-z^{\ast}\right)^2|s|^2,
\end{equation}
which is negative for any $z\neq z^{\ast}$ whenever $B<0$, a condition that is guaranteed for $\kappa<\sqrt{3}\,\omega_c$. Consequently, in this regime, the collective spin flows to $z^{\ast}$, and then exhibits purely azimuthal precession at $z=z^{\ast}$ generated by $H$, thereby leading to the formation of the chiral LC. Finally, when $g_1=g_2$, $z^{\ast} = z_{ss}^{\rm SR}$ and the azimuthal precession rate $\partial H[z]/\partial z$ vanishes, leading to the superradiant steady state. 

\emph{Conclusions and outlook:} In this Letter, we answered the two questions posed at the outset: dissipation does not preclude $U(1)$-broken phases of matter, and these phases can indeed be chiral. Concretely, we present the dynamical phase diagram of the open chiral Dicke model in the thermodynamic limit. We demonstrate that dissipation significantly modifies the phase diagram and the $U(1)$-broken superradiant state survives over a considerably smaller parameter regime. However, $U(1)$-breaking itself persists over a much larger portion of the phase diagram, and it is manifested in the form of two classes of LCs, distinguished by opposite chiralities. These two classes are separated by the domain-wall-like superradiant state, thereby lending a topological character to the phase diagram. Crucially, we obtain exact analytical expressions that describe the chiral LC solutions in the thermodynamic limit. Notably, these chiral LCs are robust to $U(1)$-preserving perturbations, and they serve as promising candidates for chiral time crystals. Finally, we establish that this system hosts multistable regimes, where the long-time behavior of the system is dictated by the initial state.

Several interesting avenues for future research emerge from our work. A natural first step is to explore the fate of these dynamical phases for finite $N$, where the LC oscillation is expected to persist for a time, $\tau$, that grows with $N$, and diverges when $N \rightarrow \infty$. Constructing a quantum atom-only theory which reproduces Eq.~(\ref{eq:spindyn}) in the thermodynamic limit would provide a natural route to address this question, since the resulting Hilbert space is $(N+1)$-dimensional, thereby making exact diagonalization of the Liouvillian tractable. Another direction is to investigate the dynamical phases that emerge in coupled open chiral Dicke models. Finally, it would be interesting to characterize the chiral non-equilibrium phases realized under periodic~\cite{gong2018discrete,zhu2019dicke} and quasiperiodic driving~\cite{anisur2026dissipationstabilizesdicketime} in the chiral DM.



\vspace{5mm}
\begin{acknowledgments}
This work was supported by AFOSR Grant No.~FA9550-23-1-0598 (NY and WVL). NY thanks Diego Barberena for insightful discussions. The authors acknowledge the use of Claude (Anthropic) for minor
assistance with polishing portions of the manuscript. All scientific
content, analyses, interpretations, and conclusions were developed and
verified by the authors, who take full responsibility for the manuscript.
\end{acknowledgments}

\bibliography{references}

\onecolumngrid
\begin{center}
\newpage
\textbf{
Supplemental Material:\\[4mm]
\large Topological chirality of dissipative limit cycles in an open Dicke model \\ }

\vspace{4mm}
Nikolay Yegovtsev,$^{1,2}$ Sayan Choudhury,$^{3,4}$ W. Vincent Liu$^1$ \\
\vspace{2mm}
{\em \small
$^1$Department of Physics and Astronomy and IQ Initiative, University of Pittsburgh, Pittsburgh, Pennsylvania 15260, USA \\
$^2$Department of Physics and Astronomy, Purdue University, West Lafayette, Indiana 47907, USA\\
$^3$Harish-Chandra Research Institute, a CI of Homi Bhabha National Institute, Chhatnag Road, Jhunsi, Prayagraj 211 019\\
$^4$Homi Bhabha National Institute, Training School Complex,
Anushakti Nagar, Mumbai 400 094, India
}
\end{center} 

\setcounter{equation}{0}
\setcounter{figure}{0}
\setcounter{table}{0}
\setcounter{section}{0}
\setcounter{page}{1}
\makeatletter
\renewcommand{\theequation}{S.\arabic{equation}}
\renewcommand{\thefigure}{S\arabic{figure}}
\renewcommand{\thetable}{S\arabic{table}}
\renewcommand{\thesection}{S.\arabic{section}}
\renewcommand{\theHequation}{S.\arabic{equation}}
\renewcommand{\theHfigure}{S\arabic{figure}}
\renewcommand{\theHtable}{S\arabic{table}}
\renewcommand{\theHsection}{S.\arabic{section}}

\section{Linear stability analysis of the steady states and limit cycles}
In the main text we obtained the expressions for the two trivial steady states and the superradiant state. Now we write $\alpha_1 = \alpha_{1ss}+\delta{\alpha_1}$ and similarly for other variables and study the linearized dynamics originating from Eq.~\eqref{eq:eom}:
\begin{equation}
\label{eq:lsa}
\begin{split}
&\delta\dot{\alpha}_1 = -(\kappa+i\omega_c)\delta\alpha_1 -\frac{ig_1}{\sqrt{N}}\delta s, \\
&\delta\dot{\alpha}_2 = -(\kappa+i\omega_c)\delta\alpha_2 -\frac{ig_2}{\sqrt{N}}\delta s^*, \\
&\delta\dot{s} = -i\omega_z\delta s +\frac{2i}{\sqrt{N}}z_{ss}(g_1\delta\alpha_1+g_2\delta\alpha_2^*)+\frac{2i}{\sqrt{N}}(g_1\alpha_{1ss}+g_2\alpha_{2ss}^*)\delta z,\\
&\delta\dot{z} = \frac{ig_1}{\sqrt{N}}(-\alpha_{1ss}\delta s^*-s_{ss}^*\delta\alpha_1+\alpha_{1ss}^*\delta s+s_{ss}\delta\alpha_1^*) + \frac{ig_2}{\sqrt{N}}(\alpha_{2ss}\delta s + s_{ss}\delta\alpha_{2} - \alpha_{2ss}^*\delta s^*-s_{ss}^*\delta\alpha_2^*).    
\end{split}    
\end{equation}
Solving Eq.~\eqref{eq:lsa} for the normal and inverted states: $\alpha_{1ss}=\alpha_{2ss} = s_{ss} = 0$ and $z_{ss} = \mp N/2$, the eigenvalues can be obtained by solving the polynomial equation:
\begin{equation}
\begin{split}
&\omega^6 +4\kappa\omega^5+\left[\omega_z^2+2\left(3\kappa^2+\omega_c^2\pm(g_1^2-g_2^2)\right)\right]\omega^4 +2\kappa\left[2(\kappa^2+\omega_c^2+\omega_z^2)\pm3(g_1^2-g_2^2)\right]\omega^3\\
&+\left[(g_1^2-g_2^2)^2+(\kappa^2+\omega_c^2)^2+2\omega_z^2(3\kappa^2+\omega_c^2)\pm2(3\kappa^2+\omega_c^2)(g_1^2-g_2^2)\mp2\omega_z\omega_c(g_1^2+g_2^2)\right]\omega^2\\
&+2\kappa\left[(g_1^2-g_2^2)^2\pm (\kappa^2+\omega_c^2)(g_1^2-g_2^2)\mp2\omega_z\omega_c(g_1^2+g_2^2)+2\omega_z^2(\kappa^2+\omega_c^2)\right]\omega\\
&+(\kappa^2+\omega_c^2)\left[(g_1^2-g_2^2)^2\mp2\omega_z\omega_c(g_1^2+g_2^2)+\omega_z^2(\kappa^2+\omega_c^2)\right]+4\omega_c^2g_1^2g_2^2=0.
\end{split}    
\end{equation}
For a given choice of parameters $\omega_c$, $\omega_z$ and $\kappa$, the regions of stability of the corresponding steady states in the $(g_1,g_2)$ plane are found by setting the real part of all eigenvalues to be negative.

For the superradiant state, we set $g_1=g_2=g$ and use expressions in Eq.~\eqref{eq:zss} and Eq.~\eqref{eq:sr} of the main text. This results in the following eigenvalue equation:
\begin{equation}
\begin{split}
&\omega^7 +4\kappa\omega^6 + \left(2 \omega_c ^2
   \left(\frac{2 g^4}{\left(\kappa
   ^2+\omega_c ^2\right)^2}+1\right)+6
   \kappa ^2\right)\omega ^5 +\frac{4 \kappa   \left(4 g^4
   \omega_c ^2+\left(\kappa ^2+\omega_c
   ^2\right)^3\right)}{\left(\kappa
   ^2+\omega_c ^2\right)^2}\omega ^4   \\
&+\frac{ \left(20 g^4 \kappa ^2
   \omega_c ^2+4 g^4 \omega_c ^4+\kappa ^8+4
   \kappa ^6 \omega_c ^2+6 \kappa ^4 \omega_c
   ^4+4 \kappa ^2 \omega_c ^6-\omega_z^2 \left(\kappa ^2+\omega_c
   ^2\right)^3+\omega_c
   ^8\right)}{\left(\kappa ^2+\omega_c
   ^2\right)^2}\omega ^3 \\
& +  \left(-2 \kappa  \omega_z^2 \left(\kappa ^2+\omega_c
   ^2\right)+\frac{8 g^4 \kappa  \omega_c
   ^2}{\kappa ^2+\omega_c ^2}\right)\omega ^2 = 0.
\end{split}    
\end{equation}
We observe that this equation has two eigenvalues $\omega = 0$. The first one corresponds to the conservation of the total spin length, and the other to the arbitrary choice of $\phi$ in the symmetry broken phase. The last term in the above equation vanishes at the superradiant transition $g = \sqrt{\omega_z(\kappa^2+\omega_c^2)/(2\omega_c)}$. All eigenvalues have negative real part above that point, so the superradiant phase is stable.

The stability of limit cycles can be checked using a similar method. For a given point in the LC phase diagram, we linearize the equations of motion, and this results in a system of equations that has the same form as Eq.~\eqref{eq:lsa}, where instead of steady-state solutions, we substitute the solutions corresponding to limit cycles. The stability of the system can be inferred from the Floquet multipliers, which must all lie inside the unit circle on the complex eigenvalue plane.

\section{Atom-only theory in the semiclassical limit}
\subsection{Standard Dicke model}
Consider the standard Dicke Hamiltonian:
\begin{equation}
H = \omega_c a^\dagger a +\omega_zS^z +\frac{2g}{\sqrt{N}}(a^\dagger+a)S^x.    
\end{equation}
The semiclassical equations of motion for $\alpha = \langle a\rangle$, $x = \langle S^x\rangle$, $y = \langle S^y\rangle$, $z = \langle S^z\rangle$ are:
\begin{equation}
\begin{split}
&\dot{\alpha} = -(\kappa+i \omega_c)\alpha - \frac{2ig}{\sqrt{N}}x,\\
&\dot{x} = -\omega_z y,\\
&\dot{y} = \omega_z x-\frac{2g}{\sqrt{N}}(\alpha+\alpha^*)z,\\
&\dot{z} = \frac{2g}{\sqrt{N}}(\alpha+\alpha^*)y. 
\end{split}  
\end{equation}
Next, we write $\alpha = \alpha^{(0)}+\alpha^{(1)}$, where $\alpha^{(0)}$ is the expression obtained by standard adiabatic elimination:
\begin{equation}
\label{eq:alpha0}
\alpha^{(0)} = -\frac{2ig}{\sqrt{N}(\kappa+i\omega_c)}x,
\end{equation}
while $\alpha^{(1)}$ captures the effect of dissipation and relaxes the system to the corresponding steady state. By plugging this back into the equation for $\dot{\alpha}$, we get:
\begin{equation}
\label{eq:alpha1}
\alpha^{(1)} = -\frac{2i\omega_z g}{\sqrt{N}(\kappa+i\omega_c)^2}y.    
\end{equation}
Finally, plugging the result for $\alpha$ into the remaining equations, we get:
\begin{equation}
\label{eq:DM}
\begin{split}
&\dot{x} = -\omega_z y,\\
&\dot{y} = \omega_z x+\frac{8\omega_c g^2}{N(\kappa^2+\omega_c^2)}xz+\frac{16\omega_z\omega_c\kappa g^2}{N(\kappa^2+\omega_c^2)^2}yz,\\
&\dot{z} = -\frac{8\omega_c g^2}{N(\kappa^2+\omega_c^2)}xy - \frac{16\omega_z\omega_c \kappa g^2}{N(\kappa^2+\omega_c^2)^2}y^2.
\end{split}    
\end{equation}
This dynamics can be thought of as arising from the classical spin dynamics obeying equations:
\begin{equation}
\dot{\mathbf{S}} = \{\mathbf{S},H\} +\Gamma_2\mathbf{S}\times \hat{x},    
\end{equation}
with the Hamiltonian:
\begin{equation}
H = \omega_zS^z - \frac{4\omega_cg^2}{N(\kappa^2+\omega_c^2)}(S^x)^2,    
\end{equation}
and the spin-dependent damping term:
\begin{equation}
\Gamma_2 = \frac{16\omega_z\omega_c \kappa g^2}{N(\kappa^2+\omega_c^2)^2}S^y    
\end{equation}

This reproduces the results in equations (35,36) of \cite{atomonlyz2}. One can derive an effective quantum model that generates the same equations in the semiclassical limit. One needs to substitute the expression for $\alpha$ back into the Hamiltonian and neglect the terms that are either small in $1/N$ or are of higher order in the expansion parameter $\omega_z/(\kappa^2+\omega_c^2)$ compared to the ones written in Eq.~\eqref{eq:DM}. The resultant dynamics can be expressed in the Lindbladian form $\dot{\rho} = -i[H,\rho] + \kappa\left(2L^\dagger \rho L - \{L^\dagger L, \rho\}\right)$ with the atom-only Hamiltonian:
\begin{equation}
H_{atoms} = \omega_z S^z - \frac{4\omega_c g^2}{N(\kappa^2+\omega_c^2)}(S^x)^2-\frac{4\kappa \omega_c \omega_z g^2}{N(\kappa^2+\omega_c^2)^2}\left(S^xS^y+S^yS^x\right),      
\end{equation}
and the jump operator $L = \tilde{\alpha}^{(0)}S^x+\tilde{\alpha}^{(1)}S^y$, where $\tilde{\alpha}^{(0)}$ and $\tilde{\alpha}^{(1)}$ are given by the coefficient in front of $x$ and $y$ in Eq.~\eqref{eq:alpha0} and Eq.~\eqref{eq:alpha1}

\subsection{Chiral Dicke model}
Here we show how to generalize the above method to the $U(1)$ case. The Hamiltonian is:
\begin{equation}
H_{\rm CDM} =  \omega_c(a_1^\dagger a_1+a_2^\dagger a_2)+\omega_zS^{z} +\frac{g_1}{\sqrt{N}}(a_1S^{+}+a_1^\dagger S^{-}) +\frac{g_2}{\sqrt{N}}(a_2S^{-}+a_2^\dagger S^{+}).    
\end{equation}
The mean-field equations:
\begin{equation}
\label{eq:eom}
\begin{split}
&\dot{\alpha}_1 = -(\kappa+i\omega_c)\alpha_1- \frac{ig_1}{\sqrt{N}}s,\\
&\dot{\alpha}_2 = -(\kappa+i\omega_c)\alpha_2- \frac{ig_2}{\sqrt{N}}s^*,\\
&\dot{s} = -i\omega_zs +\frac{2i}{\sqrt{N}}(g_1\alpha_1+g_2\alpha_2^*)z,\\
&\dot{z} = \frac{ig_1}{\sqrt{N}}\left(-\alpha_1s^*+\alpha_1^* s\right)+\frac{ig_2}{\sqrt{N}}\left(\alpha_2s-\alpha_2^* s^*\right).
\end{split}    
\end{equation}
Next, we write $\alpha_1 = \alpha_1^{(0)}+\alpha_{1}^{(1)}$ and $\alpha_2 = \alpha_2^{(0)}+\alpha_{2}^{(1)}$, where $\alpha_{1}^{(0)}$ and $\alpha_2^{(0)}$ are the expressions obtained by standard adiabatic elimination:
\begin{equation}
\begin{split}
&\alpha_1^{(0)} = -\frac{ig_1}{\sqrt{N}(\kappa+i\omega_c)}s,\\
&\alpha_2^{(0)} = -\frac{ig_2}{\sqrt{N}(\kappa+i\omega_c)}s^*,
\end{split}    
\end{equation}
while $\alpha_1^{(1)}$ and $\alpha_{2}^{(1)}$ capture the effects of dissipation and relax the system to the corresponding state. By plugging these expressions back into the mean-field equations and using the method of successive approximations, we get:
\begin{equation}
\begin{split}
&\alpha_1^{(1)} = \frac{ig_1}{\sqrt{N}(\kappa+i\omega_c)^2}\left(-i\omega_z s+\frac{2i}{\sqrt{N}}(g_1\alpha_1^{(0)}+g_2\alpha_2^{(0)*})z\right),\\
&\alpha_2^{(1)} = \frac{ig_2}{\sqrt{N}(\kappa+i\omega_c)^2}\left(i\omega_z s^*-\frac{2i}{\sqrt{N}}(g_1\alpha_1^{(0)*}+g_2\alpha_2^{(0)})z\right).
\end{split}    
\end{equation}
Plugging the above results for $\alpha_1$ and $\alpha_2$ into the equations for $\dot{s}$, $\dot{z}$, we finally arrive at the atom-only description:
\begin{equation}
\begin{split}
&\dot{s} = -i\omega_zs -\frac{2\kappa(g_2^2-g_1^2)+2i\omega_c(g_1^2+g_2^2)}{N(\kappa^2+\omega_c^2)}sz+\frac{4\kappa\omega_c\omega_z(g_1^2+g_2^2)-2i(\kappa^2-\omega_c^2)\omega_z(g_2^2-g_1^2)}{N(\kappa^2+\omega_c^2)^2}sz\\&-\frac{4\left[(\kappa-i\omega_c)^3g_1^4+(\kappa+i\omega_c)^3g_2^4-2\kappa(\kappa^2+\omega_c^2) g_1^2g_2^2\right]}{N^2(\kappa^2+\omega_c^2)^3}sz^2,\\
&\dot{z} = \frac{2\kappa(g_2^2-g_1^2)}{N(\kappa^2+\omega_c^2)}|s|^2-\frac{4\kappa\omega_c\omega_z(g_1^2+g_2^2)}{N(\kappa^2+\omega_c^2)^2}|s|^2+\frac{4\kappa^3(g_1^2-g_2^2)^2-4\kappa\omega_c^2(3g_1^4+3g_2^4+2g_1^2g_2^2)}{N^2(\kappa^2+\omega_c^2)^3}|s|^2z
\end{split}    
\end{equation}
We note that the above equations are suitable to describe both the steady states as well as the limit-cycle phases. The resultant theory still has a $U(1)$ symmetry associated with the rotations $s\to e^{i\phi}s$. The above equations of motion can be also thought of as arising from the:
\begin{equation}
\dot{\mathbf{S}} = \{\mathbf{S},H\} - \Gamma_1\mathbf{S}\times(\mathbf{S}\times \hat{z}),       
\end{equation}
with the LMG-like Hamiltonian:
\begin{equation}
H = \omega_zS^z - \frac{4\omega_c(\kappa^2-\omega_c^2)(g_2^4-g_1^4)}{3N^2(\kappa^2+\omega_c^2)^3}(S^z)^3- \left[\frac{\omega_c(g_1^2+g_2^2)}{N(\kappa^2+\omega_c^2)}-\frac{\omega_z(\kappa^2-\omega_c^2)(g_2^2-g_1^2)}{N(\kappa^2+\omega_c^2)^2}\right]\left[(S^x)^2+(S^y)^2\right],    
\end{equation}
and dissipative term takes the form:
\begin{equation}
    \Gamma_1 = A + B S_z, 
\end{equation}
where
\begin{eqnarray}
A&=& \frac{2\kappa(g_2^2-g_1^2)}{N(\kappa^2+\omega_c^2)}-\frac{4\kappa\omega_c\omega_z(g_1^2+g_2^2)}{N(\kappa^2+\omega_c^2)^2} \nonumber\\
B&=&\frac{4\kappa\left[\kappa^2(g_1^2-g_2^2)^2-\omega_c^2(3g_1^4+3g_2^4+2g_1^2g_2^2)\right]}{N^2(\kappa^2+\omega_c^2)^3}      
\end{eqnarray}

\section{Robustness of the phase diagram}
\subsection{Robustness towards the change of parameters}
Let us consider the case when each cavity mode has different frequency and dissipation rate:
\begin{equation}
\begin{split}
&\dot{\alpha}_1 = -(\kappa_1+i\omega_{1})\alpha_1-\frac{ig_1}{\sqrt{N}}s,\\    
&\dot{\alpha}_2 = -(\kappa_2+i\omega_{2})\alpha_2-\frac{ig_2}{\sqrt{N}}s^*,\\
&\dot{s} = -i\omega_zs +\frac{2i}{\sqrt{N}}(g_1\alpha_1+g_2\alpha_2^*)z,\\
&\dot{z} = \frac{ig_1}{\sqrt{N}}\left(-\alpha_1s^*+\alpha_1^* s\right)+\frac{ig_2}{\sqrt{N}}\left(\alpha_2s-\alpha_2^* s^*\right).
\end{split}    
\end{equation}
By looking for the limit cycle solution in the same form as in the main text, we get the condition on the coupling:
\begin{equation}
\label{eq:SMclc}
\frac{g_1^2\kappa_1}{\kappa_1^2+(\omega-\omega_{1})^2}-\frac{g_2^2\kappa_2}{\kappa_2^2+(\omega+\omega_{2})^2}=0.    
\end{equation}
Solution reads:
\begin{equation}
\label{eq:SMomega}
\omega = \frac{-(\kappa_1g_1^2\omega_2+\kappa_2g_2^2\omega_1)+\sqrt{(\kappa_1g_1^2\omega_2+\kappa_2g_2^2\omega_1)^2-(\kappa_1g_1^2-\kappa_2g_2^2)\left[\kappa_1g_1^2(\kappa_2^2+\omega_2^2)-\kappa_2g_2^2(\kappa_1^2+\omega_1^2)\right]}}{\kappa_1g_1^2-\kappa_2g_2^2}    
\end{equation}
The steady state superradiant phase corresponds to $\omega=0$, which gives the equation of a line that separates chiral limit cycle phases:
\begin{equation}
g_2 = \sqrt{\frac{(\kappa_2^2+\omega_{c2}^2)\kappa_1}{(\kappa_1^2+\omega_{c1}^2)\kappa_2}}g_1.    
\end{equation}
If we are slightly below this line, then we will be in $CLC_{-}$ phase and if above in the $CLC_{+}$ phase just like in Fig.\ref{fig:1} of the main text. This allows us to define chirality as:
\begin{equation}
C = \text{sign}\left(g_2-\sqrt{\frac{\kappa_1(\kappa_2^2+\omega_2^2)}{\kappa_2(\kappa_1^2+\omega_1^2)}}g_1\right).     
\end{equation}

\subsection{Robustness towards symmetry preserving terms}
Now we consider robustness of the LC phases to $U(1)$ symmetry-preserving perturbations that can arise in realistic experiments. One of such terms is $U_1S^z(a_1^\dagger a_1+a_2^\dagger a_2)/N$ and another is $U_2S^z(a_1^\dagger a_1-a_2^\dagger a_2)/N$. They can be combined into a single term of the form:
\begin{equation}
H_{pt} = S^z\left(\frac{V_1}{N}a_1^\dagger a_1+\frac{V_2}{N}a_2^\dagger a_2\right),    
\end{equation}
with $V_1 = U_1+U_2$ and $V_2=U_1-U_2$. The semiclassical equations of motion read:
\begin{equation}
\begin{split}
&\dot{\alpha}_1 = -\left(\kappa+i\left[\omega_c+\frac{V_1}{N}z\right]\right)\alpha_1-\frac{ig_1}{\sqrt{N}}s,\\
&\dot{\alpha}_2 = -\left(\kappa+i\left[\omega_c+\frac{V_2}{N}z\right]\right)\alpha_2-\frac{ig_2}{\sqrt{N}}s^*,\\
&\dot{s} = -i\left(\omega_z+\frac{V_1}{N}|\alpha_1|^2+\frac{V_2}{N}|\alpha_2|^2\right)s+\frac{2i}{\sqrt{N}}(g_1\alpha_1+g_2\alpha_2^*)z,\\
&\dot{z} = \frac{ig_1}{\sqrt{N}}\left(-\alpha_1s^*+\alpha_1^* s\right)+\frac{ig_2}{\sqrt{N}}\left(\alpha_2s-\alpha_2^* s^*\right).
\end{split}
\end{equation}
Again, using the ansatz for the LC solutions, we get:
\begin{equation}
\begin{split}
&\alpha_1(t) = \frac{\sqrt{N}g_1\sin(\theta)}{2}\frac{e^{-i\omega t +i\phi}}{(\omega-\left[\omega_c+\frac{V_1}{2}\cos(\theta)\right])+i\kappa},\\
&\alpha_2(t) = -\frac{\sqrt{N}g_2\sin(\theta)}{2}\frac{e^{i\omega t -i\phi}}{(\omega+\left[\omega_c+\frac{V_2}{2}\cos(\theta)\right])-i\kappa}.
\end{split}    
\end{equation}
Plugging into equation for $\dot{z}$, we get:
\begin{equation}
\frac{g_1^2}{\left(\omega-\omega_c-\frac{V_1}{2}\cos(\theta)\right)^2+\kappa^2} - \frac{g_2^2}{\left(\omega+\omega_c+\frac{V_2}{2}\cos(\theta)\right)^2+\kappa^2}=0 
\end{equation}
Note that unlike previous cases discussed so far, we cannot extract $\omega$ directly, and also need to consider $\dot{s}$ as well to find explicit expressions for $\omega$ and $\theta$. This approach is not very tractable, so we do not pursue it further. Note, however, that $\omega_1=\omega_c+V_1\cos(\theta)/2$ and $\omega_2=\omega_c+V_2\cos(\theta)/2$ can be interpreted as having two cavity modes with different frequencies, so provided we know $\theta$, we can readily use our previous result in Eq.~\eqref{eq:SMomega}.
\end{document}